# Kossel diffraction and photonic modes in one-dimensional photonic crystal

J.-M. André[1,2], P. Jonnard[1,2], K. Le Guen[1,2] and F. Bridou[3]

[1] *Sorbonne Universités, UPMC Univ Paris 06, Laboratoire de Chimie Physique-Matière et Rayonnement, 11 rue Pierre et Marie Curie, F-75231 Paris cedex 05, France*

[2] *CNRS UMR 7614, Laboratoire de Chimie Physique-Matière et Rayonnement, 11 rue Pierre et Marie Curie, F-75231 Paris cedex 05, France*

[3] *Laboratoire Charles Fabry, Institut d'Optique, CNRS UMR 8501, Université Paris Sud 11, 2 avenue Augustin Fresnel, F-91127 Palaiseau Cedex, France*

E-mail: jean-michel.andre1@upmc.fr

**Abstract**

Kossel diffraction under standing-wave excitation in a one-dimensional photonic crystal is investigated. It is shown that by combining the reciprocity theorem, the Fermi golden rule and the concept of density of photonic modes, it is possible to predict the behaviour of the Kossel diffraction in such a system.

## 1. Introduction

Standard Kossel lines, that is the spatial distribution of the x-ray fluorescence emission from a crystal induced by ionizing radiation were first recorded in the thirties [1]. The shape of the lines was explained by using both the dynamical theory of diffraction and the reciprocity theorem [2]. The inherent equivalence of Kossel diffraction and x-ray standing-waves (XSW) was pointed out by T. Gog *et al.* [3] by virtue of the reciprocity theorem, which according to James [4], states that "if a source of radiation and a point of observation are interchanged, the electric field intensity will be the same in the new point of observation as it was at the old". Accordingly, Kossel diffraction and standing-wave mechanism can be viewed as space reversed processes : the fluorescing atoms within the crystal are sources of radiation that can be observed by a distant detector. Then it becomes clear that Kossel diffraction and XSW can be implemented and combined for structural characterization of crystal. This technique became powerful with the advent of synchrotron radiation facilities [5].

Kossel diffraction and standing-wave mechanism has been extended to the new artificial crystals, the so-called photonic crystals [6,7]. In particular, Kossel lines from the one-dimensional photonic crystals (1D-PCs) with nanometer scale developed for x-ray optics, that is the Bragg multilayer reflectors, were observed under electron [8,9] or x-ray [10–13] excitation. The analysis of a periodic structure stratified on a nanometer scale using fluorescence modulated by a XSW, that can be regarded as a Kossel method, is a rapidly developing method which makes it possible to investigate the atomic distribution especially at the interfaces [14].

In this paper we study the Kossel diffraction under standing-wave excitation in a one-dimensional photonic crystal. We will show that using the reciprocity theorem, the Fermi golden rule and the concept of density of photonic modes, it is possible to predict the behaviour of the Kossel diffraction in such a system. We illustrate our approach by analyzing the Kossel diffraction by a Fe/C Bragg mirror under XSW excitation at 8 keV.

## 2. Application of the reciprocity theorem to the Kossel effect in a 1D-PC

### 2.1 First method approach

We consider a 1D-PC as shown in Figure 1 formed by a periodic stack of N bilayers. The element of one of the layers (say 1, or Fe in the case described below) can emit x-ray fluorescence of photon energy $E_{fluo}$ upon excitation by monochromatic radiation of energy $E_{exc}$ in the geometry sketched in Fig. 1.

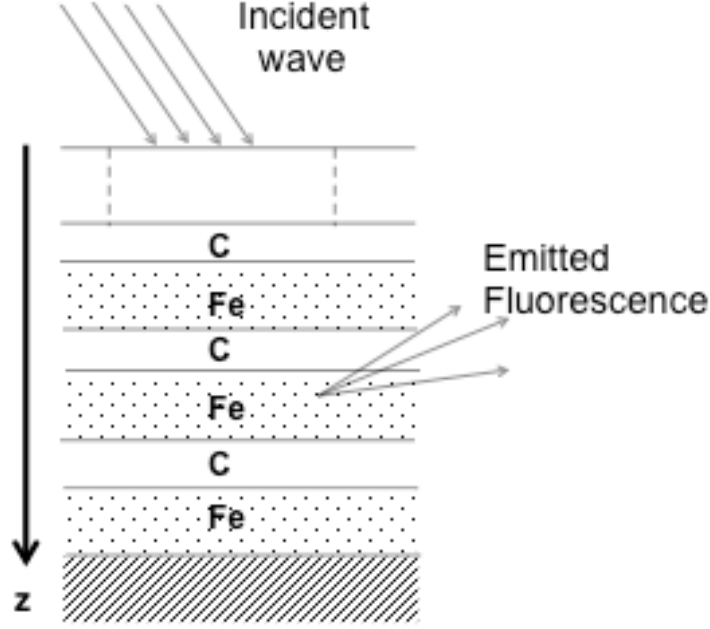

Figure 1: Scheme of the one-dimensional crystal formed by a periodic stack of N bilayers. The bilayer is made up with a medium 1, Fe in our case, emitting fluorescence and a medium 2, C in our case.

The intensity $I_{fluo}$ of the fluorescence radiation at the depth $z$ of the structure is given by the formula:

$$I_{fluo}(z, \theta_{in}, \theta_{out}) = c(z)|\mathcal{E}_{exc}(z, \theta_{in})|^2 \, |\mathcal{E}_{fluo}(z, \theta_{out})|^2$$

(1)

where $\mathcal{E}_{exc}(z, \theta_{in})$ stands for the electric field of the exciting (primary) radiation and $\mathcal{E}_{fluo}(z, \theta_{out})$ is the electric field of the fluorescent (secondary) radiation; $c(z)$ denotes the concentration of fluorescing atoms at $z$; $\theta_{in}$ and $\theta_{out}$ are the angles of the incident (primary) and emitted (fluorescence) radiations, respectively. This formula assumes that secondary fluorescence is negligible.

The term corresponding to the primary field $\mathcal{E}_{exc}$ is generally computed using a direct approach implementing different methods : recursive (Parratt) method [15], transfer matrix or coupled-wave theory [16] while the term $\mathcal{E}_{fluo}$ corresponding to the fluorescence radiation is generally estimated on the basis of the reciprocity theorem. The distance between the source and the detector being very large with respect to the layer thickness, $\mathcal{E}_{fluo}$ is calculated by using the same formalism as the one used to calculate $\mathcal{E}_{exc}$ and assuming a virtual source located at infinity in the detection direction. In this calculation, the results are given with an unknown proportional coefficient. Let us remark that this method is relatively time-consuming : since x-ray fluorescence is a volume scattering effect within the multilayer, it is necessary to calculate the fields at discrete equidistant virtual layers into the multilayer (meshing approach), before summing the contributions [17].

2.2 Second method approach

As we show in this section, it is very instructive and also efficient from the computational point of view to use an indirect approach (hereafter called second method) also using the reciprocity theorem to calculate the term $I_{exc} = |\mathcal{E}_{exc}(z, \theta_{in})|^2$. Indeed, by virtue of this theorem, as indicated previously, one can also calculate the intensity $I_{exc}$ of the exciting field $\mathcal{E}_{exc}(z, \theta_{in})$, as the one of the electric field generated in the far-field by radiating sources properly distributed in the PC. This intensity is proportional to the spontaneous emission rate of a transition between the initial and the final states $i$ and $f$ (of energy $E_i$ and $E_f$ respectively) involved in the fluorescence emission. The crucial point is that this rate is simply given by the Fermi golden rule [18]:

$$I_{exc} \propto \frac{2\pi}{\hbar} |\langle f|H|i\rangle|^2 \delta(E_f - E_i) \rho(\hbar \omega_k)$$

(2)

where $H$ is the Hamiltonian of the whole system (atom in PC + field) and $\rho$ the density of photonic modes (DOM) of the PC at the frequency $\omega_k$ of the emitted photon with the wave number $k$. In fact, assuming the validity of the Wigner-Weisskopf regime, the DOM is the same classically or quantum dynamically [19], so that a mere classical approach can be used to deal with the above problem.

In this context, the radiating source is described by an oscillating dipole with moment $\boldsymbol{\mu}(\boldsymbol{R})$ at the position $\boldsymbol{R}$ and a frequency $\omega_0$, so that the steady-state rate of power emission $P$ after all transients have vanished, is given by [19]:

$$P = \pi^2 \mu^2 \omega_0^2 \sum_\sigma \int |\boldsymbol{a}_{k\sigma}(\omega_k, \boldsymbol{R}) \cdot \boldsymbol{\mu}|^2 \delta(\omega_0 - \omega_{k\sigma}) d^3k$$

(3)

where $\boldsymbol{a}_{k\sigma}(\omega_{k\sigma}, \boldsymbol{R})$ stands for the normal modes (NMs) at a given polarization mode $\sigma = s$ or $p$ ; the frequency $\omega_{k\sigma}$ is related to the wavenumber $k$ by the dispersion relation which depends on the geometry of the PC. By integrating this expression over the wavenumbers $k$, the DOM $\rho(\omega_{k\sigma}) = dk/d\omega_{k\sigma}$ appears as a result of changing the variable $k$ to $\omega_{k\sigma}$. It yields the power spectrum $P_\omega$ of the emitted radiation as follows :

$$P_\omega = C\, \rho(\omega_k)|\mathbf{a}_k(\omega_k, \mathbf{R})|^2$$

(4)

*C* is some constant irrelevant in this case. It includes includes the polarization effects : since both dipole moment and NMs are vector quantities, it is necessary to take into account their relative directions via the dot product in Eq.(3) ; this fact leads to a multiplying factor (2 for the two possible polarization degrees of freedom [19]) which can be included in the term *C*. Since the DOM is independent of the position $\mathbf{R}$, the power spectrum can be calculated by replacing the values $|\mathbf{a}_{k\sigma}(\omega_{k\sigma}, \mathbf{R})|^2$ by the averaged value:

$$\langle|\mathbf{a}_{k\sigma}(\omega_{k\sigma}, \mathbf{R})|^2\rangle = \frac{1}{D}\int_0^D \sigma(z, \theta_{in})|\mathbf{a}_{k\sigma}(\omega_{k\sigma}, z)|^2\, dz$$

(5)

taking into account that in the above geometry geometry, the distribution of the oscillating dipoles $\sigma(z, \theta_{in})$ varies only along the *z* axis (see figure 1) ; *D* is the length of the structure. This distribution can be calculated by considering that the dipoles are induced by the exciting electric field $\mathcal{E}_{exc}(z, \theta_{in})$ which depends on the incoming angle $\theta_{in}$; it should be noted that under the Bragg condition for the exciting radiation $\mathcal{E}_{exc}(z, \theta_{in})$ forms a system of standing-waves. The expression of the DOM in a finite 1D-PC was considered by Bendickson *et al.* [20] and extended to absorbing media and oblique incidence in Ref. [21]. For the sake of consistency, the calculation is summarized here. Let *t(ω)* be the complex transmission coefficient of the 1D-PC. Then the DOM $\rho(\omega)$ is given by:

$$\rho(\omega) = \frac{1}{D}\frac{y'x - x'y}{x^2 + y^2}$$

(6)

where *x* and *y* are the real and imaginary part of *t*, respectively ; the prime denotes differentiation with respect to ω. The coefficient *t* can be calculated by means of the transfer matrix [16]. The normal modes for a N-bilayered 1D-PC have been studied by André and Jonnard [22] in the s-polarization case extending the works done for the Kronig-Penney limit case [19,23]. In a general way, for both polarizations *s* and *p*, the NMs are the solutions of the Helmholtz vector equation:

$$\nabla \wedge \nabla \wedge \mathbf{a}_k(\omega_k, \mathbf{R}) - \left(\frac{\omega_k}{c}\right)^2 \varepsilon(\mathbf{R})\, \mathbf{a}_k(\omega_k, \mathbf{R}) = 0$$

(7)

with the completeness relationship

$$\int \varepsilon(\mathbf{R})\, \mathbf{a}_k *(\omega_k, \mathbf{R})\, \mathbf{a}_k(\omega_k, \mathbf{R})\, d\mathbf{R} = \delta_t\, (\mathbf{k} - \mathbf{k}')$$

(8)

where $\delta_t$ stands for the ε-transverse Dirac function [19]. Owing to the geometry of the problem, the NMs depend only on the *z* coordinate and Eq.(7) reduces to a Helmholtz scalar equation, whose scalar solutions $a_j(z)$ in the *j-th* layer are of the form:

$$a_j(z) = \exp(i\, p\, K\, d)[A_j \exp(i\, k_z(z - p\, d)) + B_j \exp(-i\, k_z(z - p\, d))]$$

(9)

where *K* is the Bloch wavenumber, $k_z$ the component of the wavevector along the *z* axis, *p* an integer corresponding to the $p^{th}$ order of Bragg diffraction and *d* the bilayer thickness. The coefficients $A_j$ and $B_j$ and the Bloch wavenumber can be calculated as shown in the Appendix for each polarization.

At this stage the power spectrum can be calculated as given by Eqs.(4,5). The main interest of this approach is to be efficient in terms of calculation time needed to predict the behaviour of the Kossel diffraction as a function of the incoming angle $\theta_{in}$ especially when the incident primary radiation is close to the Bragg condition. Indeed the calculation of the DOM by the method proposed in Refs. [20, 21] is very quick and makes it possible to predict the main feature of the phenomenon. This fact is illustrated in the following section.

2.3 First approach versus second one

Table 1 summarizes the different steps of two methods. It is difficult to quantify the advantage of the second approach with respect to the first one in terms of computation time but one can claim that the first method requires a language with compilation (the calculation are performed using FORTRAN 90) while calculations by the second method can be done with a mathematical programming language such as MATHEMATICA™. The requirement for a tight meshing in the first method gives rise to considerable time-consuming computations.

| Step | First method | Second method |
|---|---|---|
| 1 | Calculation of the field distribution within the multilayer from a real distant source with the exciting energy: Field 1 | Calculation of the DOM: Eq.(6) |
| 2 | Calculation of the field within the multilayer from a virtual source located at the detector position with the fluorescence energy: Field 2 | Calculation of the NMs: Eq.(9) and Appendix |
| 3 | Product of the intensity of Field 1 by intensity of Field 2: Eq.(1) | Average of the NMs using the source (fluorescent atoms) distribution in the multilayer |
| 4 |  | Product of averaged NM intensity by the DOM: Eq.(4) |

Table 1: Main steps of the two approaches.

### 4. Kossel line in a Fe/C1D-PC under XSW excitation

The following experiment is now considered whose details and first results have been reported in Ref. [11]. The 1D-PC consists in 24 Fe/C bilayers with Fe and C layer thickness equal to 2.8 and 2.6 nm respectively, which gives a period of 5.4 nm. It is irradiated by the Cu $K_\alpha$ line (8045 eV or 0.154 nm) at an angle $\theta_{in}$ in the domain of the Bragg condition for this radiation, that is around 0.9°. This radiation excites the Fe $K_\alpha$ fluorescence, which is recorded as a function of the angle $\theta_{out}$ (see Fig. 1 for the geometry). With the first calculation method, one obtains for this system a typical angular distribution displayed in Figure 2 which is in good agreement with experimental results reported in Refs. [10,11].

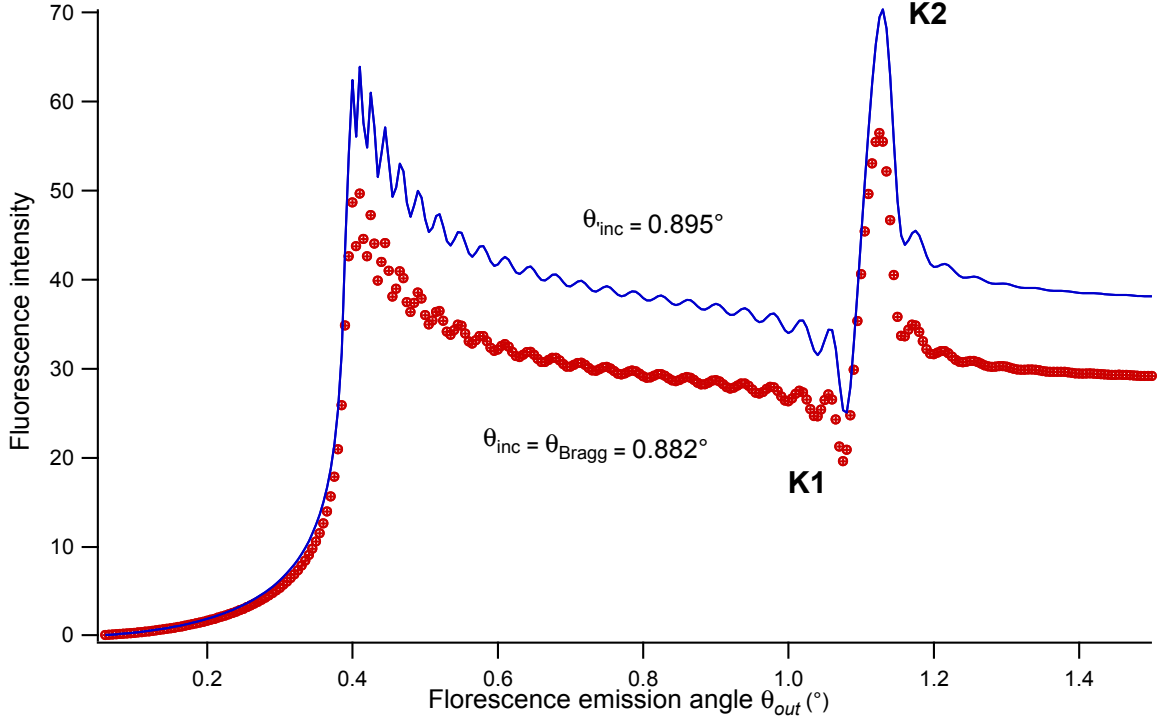

Figure 2: Calculated (first method) intensity of fluorescence radiation as a function of the emission angle $\theta_{out}$ for the Fe/C PC, for two values of the grazing incident angle: $\theta_{in} = \theta_{Bragg} = 0.882°$ (red points) and $\theta_{in} = 0.895°$ (solid blue line).

The characteristic features of a Kossel line can be seen : a dip in intensity (K1) followed by a peak in intensity (K2). The intensity of the peaks K1 and K2 varies with the incident angle $\theta_{in}$ of the primary radiation in a rather similar way. The variation results from the strong dependence of the electric field distribution intensity on the exciting Cu K$_\alpha$ radiation in the 1D-PC close to the Bragg angle. One can note that the Bragg angle (0.882°) does not give the highest value for the peak fluorescence intensity (K2), as shown in Fig. 4 : one observes that the intensity of the structure K2 is maximum around 0.91°. Indeed the Bragg angle is between the angles corresponding to K1 and K2. The features around 0.4° are related to the total reflection.

The dependence of the Kossel structures such as K1 and K2 on $\theta_{in}$ is globally well-reproduced by the variation with $\theta_{in}$ of the averaged power spectrum $\langle P_\omega(\theta_{in}) \rangle$ calculated by using Eqs.(4,5) (second method) as shown in Figure 4; it should be noted that this term reproduces (by virtue of the reciprocity theorem) the variation with the incident angle $\theta_{in}$ of the term $|\mathcal{E}_{exc}(z, \theta_{in})|^2$ and consequently of $I_{fluo}$ as given by Eq.(1). The depth-distribution of the dipole $\sigma(z, \theta_{in})$ has been computed by means of the coupled-wave theory [16]. Table 2 summarizes the main parameters used in the numerical calculations.

|  | Fe layer | C layer |
|---|---|---|
| Thickness (nm) | 2.8 | 2.6 |
| Real part of refractive index | $1 - 2.44 \, 10^{-5}$ | $1 - 7.06 \, 10^{-6}$ |
| Imaginary part of refractive index | $5.49 \, 10^{-7}$ | $1.159 \, 10^{-8}$ |

Table 2: Parameters for the calculations.

For the calculations (DOM, normal modes, …) we have considered a polarization rate equal to 50% in agreement with the experiments [10,11] which have been performed with an unpolarized source (x-ray tube). As shown in figure 3, the results given by the two approaches are in general close at least for the main features : the positions of the maxima and minima are in good agreement and the overall contrast (difference between the main minimum around 0.86° and the maximum around 0.91°) is also in agreement. In this condition we can say that the second approach is rather satisfactory. As this stage, the origin of the slight disagreement between the two approaches has not yet been accounted for.

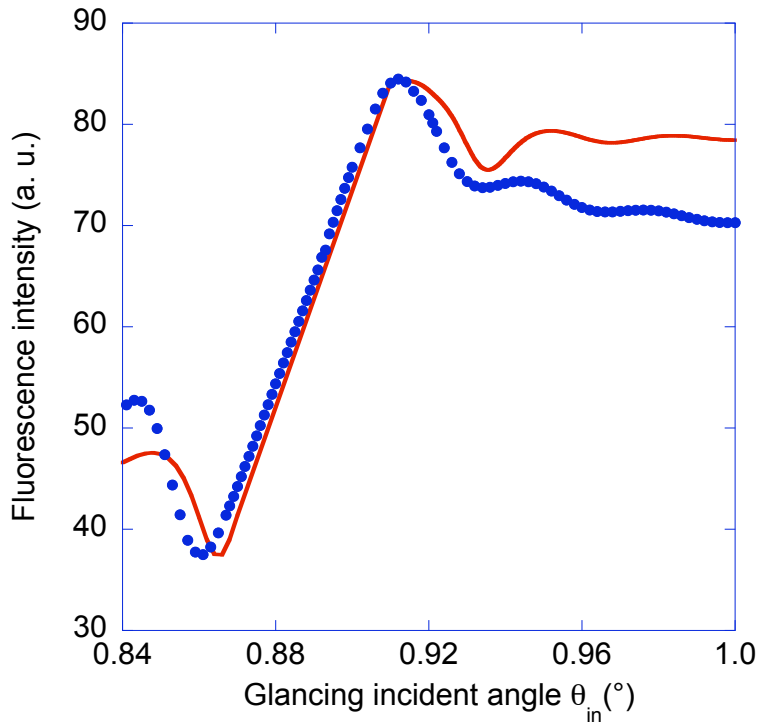

Figure 3: Variation of the fluorescence intensity of the Kossel structure K2 as a function of the incident angle $\theta_{in}$ calculated by the second method (dotted line) compared to the first method (solid line).

## 5. Conclusion and perspectives

This work shows that the combination of the reciprocity theorem, the Fermi golden rule and the concept of density of photonic modes, enables the description of the Kossel diffraction under standing-wave excitation and predicts the best conditions of illumination. This work concerning a 1D-PC could by extended to a higher dimensional PC but at the expense of a sophisticated calculation of the DOM and of the normal modes.

## APPENDIX

Let $\hat{E}_n$ be the column vector formed by the coefficients $A_n, B_n$ in Eq.(9) corresponding to the layer containing the fluorescing atoms in the $n^{th}$ bilayer :

$$\hat{E}_n = \begin{pmatrix} A_n \\ B_n \end{pmatrix}$$

(A.1)

The vector $\hat{E}_{n-1}$ corresponding to the $n$-$1^{th}$ bilayer is related to $\hat{E}_n$ by means of a matrix relationship $\bar{M}$

$$\hat{E}_{n-1} = \bar{M}_{s;p} \hat{E}_n$$

(A.2)

The matrix elements of $\bar{M}_{s;p}$ depend on the polarization case (as indicated by the subscript $s;p$) and have already been published (for instance $\bar{M}$ is given by equations (42) of reference [24]).

Moreover by virtue of the Bloch-Floquet theorem, one has

$$\hat{E}_n = e^{iKd} \hat{E}_{n-1}$$

(A.3)

where $K$ is the Bloch wavenumber.

From Eqs.(A.2) and (A.3), it follows that

$$\bar{P}_{s;p} \hat{E}_n = 0$$

(A.4)

where

$$\bar{P}_{s;p} = (\bar{M}_{s;p} - e^{iKd} \bar{I})$$

(A.5)

and $\bar{I}$ stands for the 2*2 identity matrix.

To determine all the vectors $\hat{E}_n$, it is sufficient to determine $\hat{E}_0$ and then to apply Eq.(A.2). According to Eq.(A.5), $K$ is deduced from the eigenvalues $e^{iKd}$ of the matrix $\bar{M}_{s;p}$ and the vector $\hat{E}_0$ is the eigenvector of the matrix $\bar{P}_{s;p}$ ; consequently the problem of determining $a_n(z)$ in Eq.(9) is reduced to a standard problem of linear algebra for both $s$ and $p$ polarizations.